%% file: diffusionbeam.tex
\documentclass[sponsored]{acmsiggraph}

\TOGonlineid{0077}

\TOGvolume{}
\TOGnumber{}

\TOGarticleDOI{}

\TOGprojectURL{}
\TOGvideoURL{}
\TOGdataURL{}
\TOGcodeURL{}

\title{A Dual-Beam Method-of-Images 3D Searchlight BSSRDF}

\author{ Eugene d'Eon \\
Jig Lab }

\pdfauthor{ Eugene d'Eon }


\input{smrstyle.tex}

\input{preamble.tex}

\usepackage{alltt}

\begin{document}


\newlength{\pwidth}
\newlength{\gapwidth}
\newlength{\Pwidth}

\teaser{
\centering
        \subfigure[Dipole]{\includegraphics[width=0.24 \linewidth]{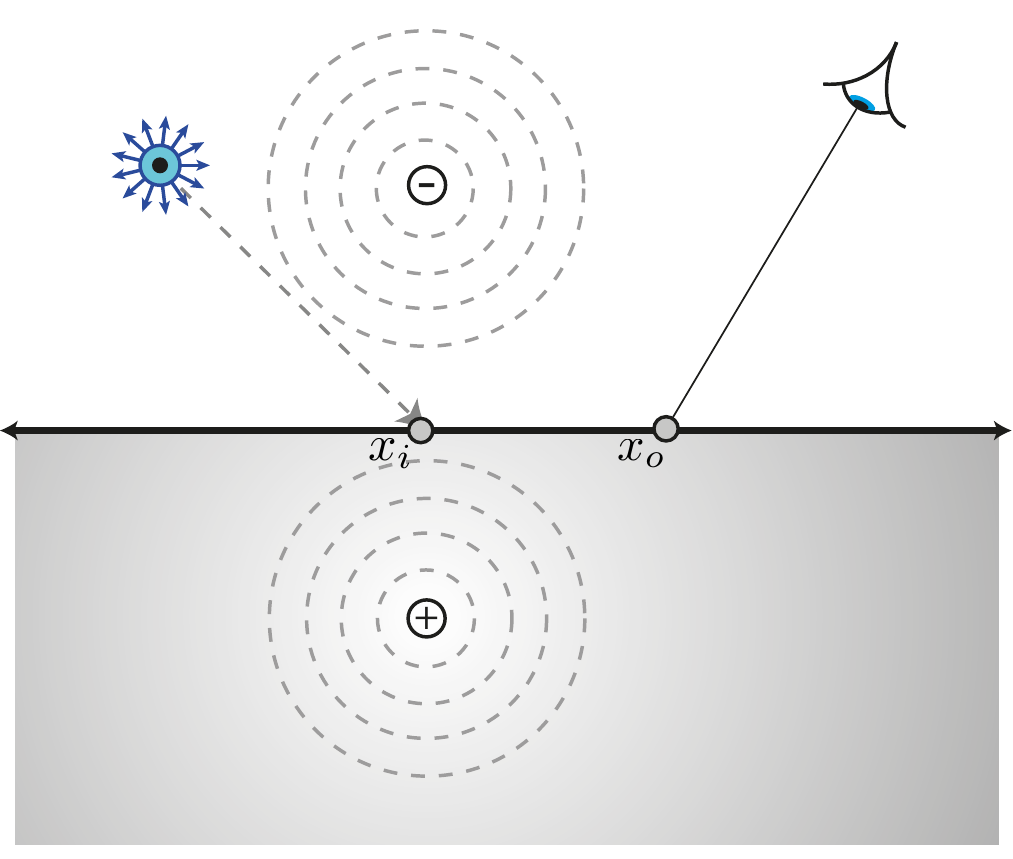}}
        \subfigure[Quantized Diffusion]{\includegraphics[width=0.24 \linewidth]{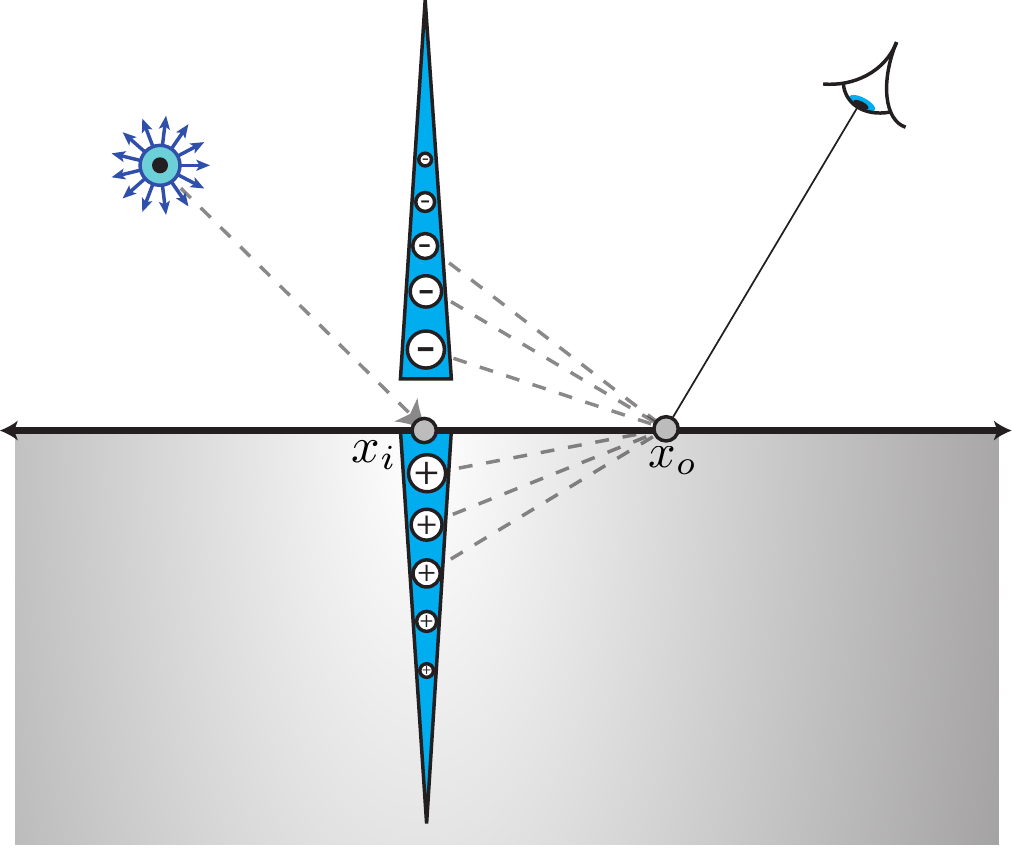}}
        \subfigure[Photon Beam Diffusion]{\includegraphics[width=0.24 \linewidth]{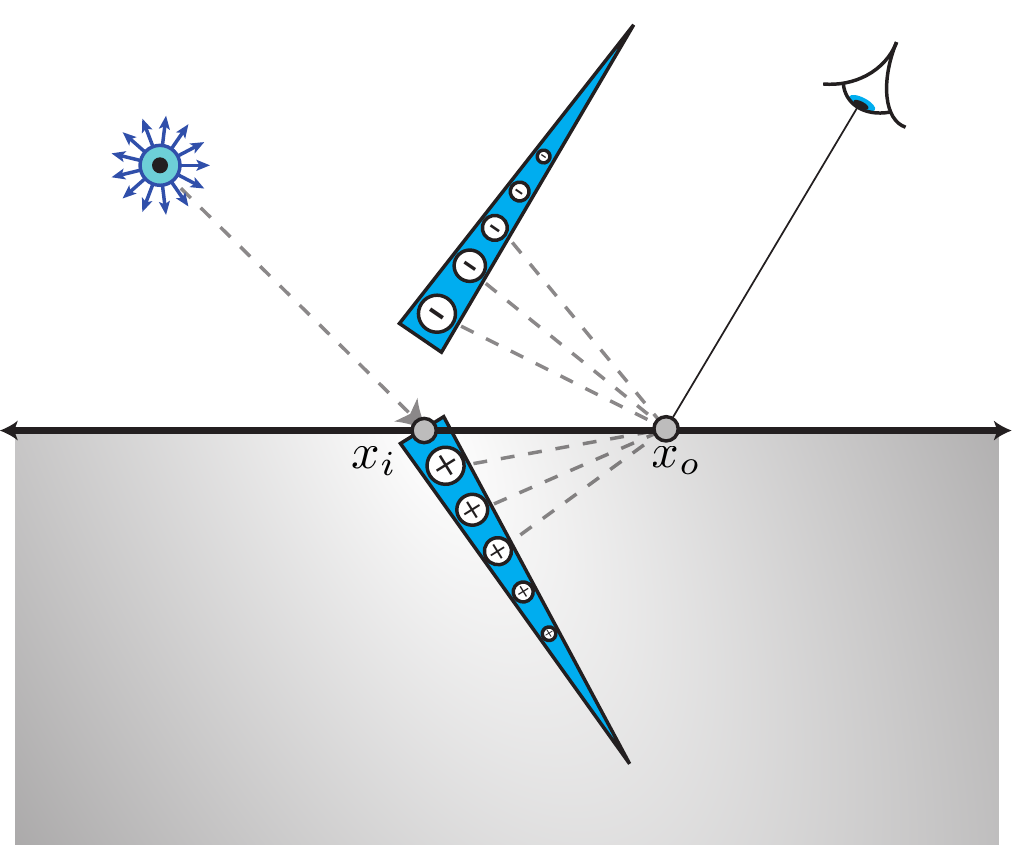}}
        \subfigure[Dual-Beam BSSRDF]{\includegraphics[width=0.24 \linewidth]{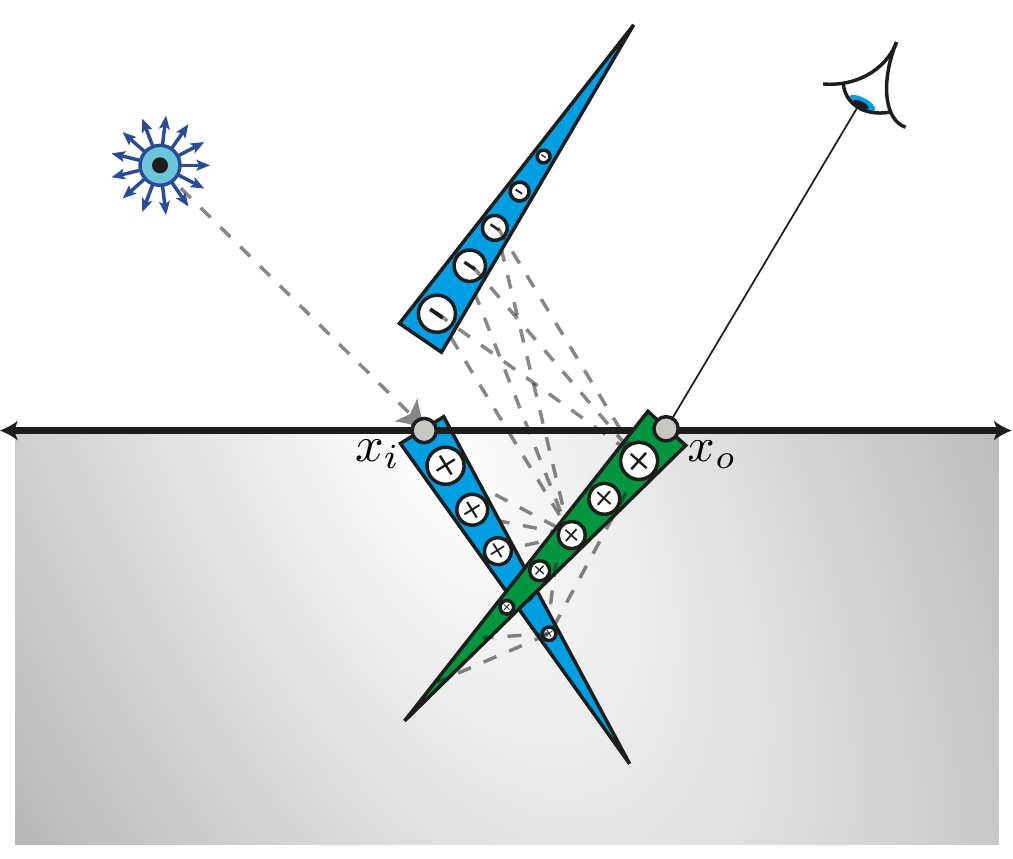}}
        \caption{The quantized-diffusion BSSRDF (b) was the first analytic BSSRDF to support exponentially-extended first-collision sources inside the translucent material, leading to improved high-frequency performance relative to the classic dipole model (a).  The photon-beam diffusion BSSRDF (c) proposes an oblique generalization of (b) for the incident refracted ray, and tilts the negative image sources outside the volume to satisfy the boundary condition.  However, this method, like (a) and (b) assumes a Fresnel-modulated Lambertian shape for the exitant radiance at $x_o$, leading to a non-reciprocal model with poor angular accuracy.  Our new dual-beam BSSRDF considers an exponentially-extended last-event detector (shown in green) inside the medium leading to the first reciprocal 8D semi-analytic BSSRDF that closely matches the associated BRDF benchmarks for the semi-infinite medium.}
        \label{fig-figs}

}


\maketitle


\begin{abstract}
  We present a novel BSSRDF for rendering translucent materials.  Angular effects lacking in previous BSSRDF models are incorporated by using a dual-beam formulation.  We employ a Placzek's Lemma interpretation of the method of images and discard diffusion theory.  Instead, we derive a plane-parallel transformation of the BSSRDF to form the associated BRDF and optimize the image confiurations such that the BRDF is close to the known analytic solutions for the associated albedo problem.  This ensures reciprocity, accurate colors, and provides an automatic level-of-detail transition for translucent objects that appear at various distances in an image.  Despite optimizing the subsurface fluence in a plane-parallel setting, we find that this also leads to fairly accurate fluence distributions throughout the volume in the original 3D searchlight problem.  Our method-of-images modifications can also improve the accuracy of previous BSSRDFs.
\end{abstract}


\begin{CRcatlist}
  \CRcat{I.3.7}{Computer Graphics}{Three-Dimensional Graphics and Realism}{Radiosity};
\end{CRcatlist}


\keywordlist






\input{intro.tex}
\input{bssrdf.tex}
\input{methodimages.tex}
\input{benchmarking.tex}

\bibliographystyle{acmsiggraph}
\bibliography{diffusionbeam}

\end{document}

%% file: smrstyle.tex
\usepackage{smrdefaults}

\usepackage{color}

\usepackage{multirow}

\usepackage{rotating}


%% file: preamble.tex
\newcommand{\vect}[1]{\vec{#1}}

\newcommand{\Gun}{G_{\text{un}}}
\newcommand{\phiun}{\phi_{\text{un}}}
\newcommand{\Gcol}{G_{\text{D}}}
\newcommand{\phicol}{\phi_{\text{D}}}
\newcommand{\Ccol}{C_{\text{D}}}
\newcommand{\mueff}{\mu_{\text{eff}}}

\newcommand{\ppdir}{u}  
\newcommand{\ssalbedo}{\alpha}

\newcommand{\arccoth}{\text{arccoth}}

\newcommand{\e}{\text{e}}

\newcommand{\x}{\vect{x}}

\newcommand{\n}{\vect{\mathbf{n}}}

\def\rm{R^{-}}

\newcommand{\textfig}[1]{\textsf{#1}}

\usepackage{paralist}
\usepackage{subfigure}
\usepackage{rotating}
\usepackage{amsgen}

%% file: intro.tex
\section{Intro}

    The importance and challenge of accurately and efficiently rendering translucent materials has led to the introduction of a number of approximate \emph{bidirectional surface-scattering reflectance distribution functions} (BSSRDFs).  For a flat, thick medium, the BSSRDF follows directly from a solution to the \emph{searchlight problem}~\cite{williams07}.  Most practical searchlight BSSRDF models in graphics~\cite{jensen01,donner05,deon11a} gain efficiency by sacrificing angular accuracy---they assume light arrives normal to the surface and derive a radial profile $R_d(r)$ for the total diffusive exitance at a given distance $r$ from the point of illumination, and this energy is spread into the outgoing directions in an ad hoc way.  In this paper we present a new BSSRDF that is related to these previous methods, but is fully 8D and is accurate in both the spatial and angular domains.  Our approach is motivated by several desired properties:
    \begin{compactitem}
        \item For use in computer graphics, the BSSRDF should be relatively simple and compact (this excludes expensive Monte Carlo methods and several recent 3D searchlight solutions~\cite{liemert12a,liemert13,gardner13})
        \item The BSSRDF should compare closely to its associated BRDF, formed by lateral integration over the surface.  This excludes non-reciprocal models that cling to a diffusion-based exitance calculation~\cite{donner07,habel13}.
    \end{compactitem}

%% file: bssrdf.tex
\section{A Dual-Beam BSSRDF}\label{sec:bssrdf}
  
    The derivation of our new BSSRDF follows from three key observations:
    \begin{compactitem}
      \item Davison~\shortcite{davison00} noted long ago that the most efficient way to compute the radiance in an isotropically-scattering \emph{infinite} medium due to an isotropic point source is by forming a line-integral of the fluence.  We employ this technique to compute the exitant radiance leaving a \emph{semi-infinite} half-space.
      \item The method of images used to formulate the diffusion dipole and related models can be interpreted as an approximate application of Placzek's Lemma~\cite{case53}---a method for constructing exact solutions to finite medium problems using only infinite medium Green's functions.  We use Grosjean's point source Green's function and revisit the method of images to better suit application of Davison's method by optimizing for accurate \emph{fluence} solutions \emph{inside} the half space instead of accurate \emph{flux at the surface}.
      \item The associated BRDF for a method of images BSSRDF can be written down in a compact form.  This is the foundation of our plane-parallel framework for efficiently applying the previous two observations to design new method-of-images BSSRDFs using the associated BRDF as an accuracy metric.
    \end{compactitem}

    We now describe our new BSSRDF.  In our notation, exitant radiance at a surface location $\x_o$ in direction $\vec{\omega_o}$ is computed by integrating the incident radiance $L_i(x_i,\vec{\omega_i})$ with a BSSRDF $S$:
    \begin{equation*}
      L_o(x_o,\vec{\omega
       }_o)=\int_{A} \int_{2\pi} S(x_i,\vec{\omega_i};x_o,\vec{\omega
       _o})L_i(x_i,\vec{\omega_i})(\vec{n} \cdot \vec{\omega_i})\, d\omega _i \, dA(x_i).
    \end{equation*}
    The BSSRDF is split into reduced-intensity, single-scattering, and multiple-scattering components, each treated separately,
    \begin{equation*}
        S = S^{(0)} + S^{(1)} + S_d.
    \end{equation*}
    For the half-space problem we consider here, $S^{(0)} = 0$.  No uncollided energy can leave the medium because we assume it is flat and infinitely thick.  Approximate application of this BSSRDF to curved geometries should compute $S^{(0)}$ using simple attenuated ray-tracing methods.  Single scattering $S^{(1)}$ can be computed using known methods~\cite{jensen01,holzschuch13}.  

    For computing the multiply-scattered portion of the BSSRDF, $S_d$, we directly apply Davison's method and write the exitant radiance as an integral of the fluence within the medium at all positions prior to their last scattering event:
    \begin{equation}\label{Sd}
      S_d\left(x_i,\vec{\omega_i};
      x_o,\vec{\omega_o}\right)
      =F_t(x_o,\vec{\omega_o},\eta) \int_0^\infty \e^{-\mu_t u} \frac{\ssalbedo}{4 \pi}\phi(\vec{x_o} + u R(\vec{\omega_o})) du.
    \end{equation}
    This integral starts at the exitant position $x_o$ and considers all locations inside the medium along the adjoint refracted ray (illustrated in green in Figure~\ref{fig-figs}d).  At each subsurface location $\vec{x}$ a portion (specifically $\ssalbedo/(4\pi)$) of the \emph{subsurface fluence} $\phi(\vec{x})$ scatters into the green outgoing refracted ray.  We must consider all locations a distance $u$ along the adjoint refracted direction $R(\vec{\omega_o})$.  At each subsurface position along this ray the energy that continues unscattering to the boundary is attenuated based on $\mu_t = \mu_s + \mu_a$, the sum of the scattering and absorption coefficients, as well as by a Fresnel transmission term $F_t$ upon exiting the medium.  The subsurface fluence $\phi$ arises due to illumination by a pencil beam striking the surface at position $\vec{x_i}$ and refracting into the medium along direction $R(\vec{\omega_i})$.  Instead of assuming all first-scatter events occur at one mean-free-path along this ray (as in the dipole model) we also form the subsurface fluence $\phi(\vec{x})$ as a line integral of first-scatter events (following Grosjean~\shortcite{grosjean58a}):
    \begin{equation}
        \phi(\vec{x}) = F_t(x_i,\vec{\omega_i},\eta) L(\vec{x_i},\vec{\omega_i}) \int_0^\infty \ssalbedo \e^{-\mu_t v} \phi_M(\vec{x}, \vec{x_i} + v R(\vec{\omega_i} )) dv.
    \end{equation}
    In contrast to previous diffusion-based BSSRDFs in graphics, here we do not compute the exitant radiance by estimating a flux balance at the boundary location $x_o$.  Instead, our new formulation requires an accurate estimation of $\phi_M(\vec{x},\vec{s})$---the fluence at position $\vec{x}$ \emph{below the surface} in a half-space due to an isotropic \emph{point source} at another subsurface location $\vec{s}$.  Thus, $\phi_M(\vec{x},\vec{s})$ is the \emph{point-to-point} Green's function for the half space.  The first interaction of the incident illumation with the medium is represented as a continuum of point sources (a beam), each of which contribute to the last-event gather beam towards the eye.  Thus, our method is a \emph{dual-beam} integral of the point-to-point Green's function.

    We approximate the point-to-point Green's function in the next section using a modified method of images (which then leads to positive and negative tilted incident beams, illustrated in Figure~\ref{fig-figs}d).  Note that $\phi_M$ excludes the reduced-intensity fluence of the incident beam itself, which includes a $\delta$ function for when the two beams cross (ignored because this corresponds to single-scattering).  

%% file: methodimages.tex
\section{The Method of Images}

    The method of images is an efficient method for constructing approximate transport theory solutions to non-infinite medium problems by linearly combining positive and negative multiples of known solutions to infinite medium problems.  As such, it is closely related to Placzek's Lemma~\cite{case53} (which proves that this process can be made exact for any convex medium).  In this section we modify the method of images for accurately approximating the fluence $\phi_M(\vec{x},\vec{s})$ at any position $\vec{x}$ within a halfspace due to a subsurface point source at position $\vec{s}$.  This contrasts from previous applications in graphics that analyze fluxes at the boundary and propose placement of negative point sources such that the exitant flux is accurate (with no specific regard to internal distributions).  Where previous methods draw upon known properties of solutions to the Milne problem for placing negative sources, we propose a novel plane-parallal framework for efficiently solving for source configurations such that desired benchmark solutions are optimally approximated, such as the associated BRDF of the half-space.

    Like the quantized-diffusion BSSRDF~\cite{deon11a}, we employ Grosjean's approximate closed-form approximation for fluence $\phi(r)$ due to an isotropic point source~\cite{grosjean56a} in an infinite isotropically-scattering medium.  To simplify the analysis, in this section we consider a homogeneous half space with isotropic scattering with single-scattering albedo $\ssalbedo$, unit interaction coefficient $\mu_t = 1$, $\mu_a = 1 - \alpha$, $\mu_s = \alpha$, and vacuum boundary conditions ($\eta = 1, F_t = 1$).

    \subsection{Infinite Medium Solutions for Plane Sources}
        We express Grosjean's infinite medium isotropic point source Green's function as a sum of uncollided (un) and diffusive (D) terms
        \begin{equation}
            G(r) = \Gun(r) + \Gcol(r) = \frac{\e^{-r}}{4 \pi r^2} + \Ccol \frac{\e^{- \mueff r}}{r}
        \end{equation}
        where
        \begin{equation}
            \Ccol = \frac{1}{4 \pi} \frac{3 \alpha}{2 - \alpha}, \,\,\, \, \, \, \, \mueff = \sqrt{\frac{\mu_a}{D}}, \,\,\, \, \, \, \, \, D = \frac{2-\alpha}{3}
        \end{equation}
        \subsubsection{The Point-to-Plane Transformation}
        The associated BRDF of a BSSRDF is found by lateral integration of the BSSRDF over all positions $x_i$ on the surface, with $\vec{\omega_i}$ and $\vec{\omega_o}$ fixed (which is then independent of $\vec{x_o}$).  This transforms isotropic point sources into isotropic plane sources.  The method of images BRDF for a half space thus requires the plane source Green's function for an infinite medium.  The exact solution is known but involves an integral over the continuous spectrum of eigenvalues of the transport operator~\cite{case67}.  Instead, we apply the plane-to-point transform~\cite{bell70} to Grosjean's approximate point source Green's function to form the \emph{laterally-integrated uncollided fluence} $\phiun$ at depth $z$ due to a plane source at depth $z_p$,
        \begin{equation}
            \phiun(z,z_p) = \int_0^\infty 2 \pi r \Gun(\sqrt{r^2 - (z-z_p)^2}) dr = -\frac{1}{2} E_i(-|z-z_p|)
        \end{equation}
        where $E_i$ is the exponential integral function~\cite{case67}.  The same transformation of the diffusive term leads to the \emph{laterally-integrated diffusive fluence} $\phicol$ at depth $z$ due to a plane source at depth $z_p$,
        \begin{equation}
            \phicol(z,z_p) = \int_0^\infty 2 \pi r \Gcol(\sqrt{r^2 - (z-z_p)^2}) dr = \frac{2 \pi \, \Ccol}{\mueff} \e^{-\mueff |z-z_p|}.
        \end{equation}
        The sum $\phiun + \phicol$ provides a useful and accurate approximation for the plane-source Green's function for an infinite medium.
%
%
    \subsection{Half-Space Solutions for plane sources}
        Applying the plane-parallel analog of the diffusion dipole, the fluence $\phi_M$ at some position in a half-space is approximated as the sum of two infinite plane-source Green's functions, one positive function at the location of the source, and one negative source mirrored about some plane, typically outside the medium.  Here, we generalize this process by separating the location of the negative uncollided and diffusive sources, as well as introducing scaling factors for each.  We suppose that the subsurface fluence is well approximated by:
        \begin{equation}
            \phi_M(\vec{x},\vec{s}) = G(||\vec{x}-\vec{s}||) - a_{un} \Gun(||\vec{x}-\vec{s_{vun}}||) - a_D \Gcol( ||\vec{x}-\vec{s_{vD}}|| )
        \end{equation}
        where position $\vec{s_{vun}}$ is the virtual source location for the uncollided portion of the Green's function, and $\vec{s_{vD}}$ is the virtual source location for the diffusive portion of the Green's function.  Placzek's lemma would seem to imply that a negative uncollided term outside the media is needed (and is consistent with previous diffusion model's overestimation of exitance).  However, the plane about which to mirror the point source is not necessarily the same for the uncollided vs the collided term, so we independently optimize for two different mirror distances, $z_{bun}$ for the uncollided portion, and $z_{bD}$ for the diffusive portion of the fluence.

        We can now write down the associated BRDF for this method-of-images BSSRDF by computing the two integrations outlined in Section~\ref{sec:bssrdf} to the plane-source transformed fluence function $\phi_M$.  We later optimize for parameters $z_{bun}$, $z_{bD}$, $a_{un}$, and $a_D$ that make this BRDF (and its related BSSRDF) accurate.  

        The component of the BRDF due to the positive uncollided term in $\phi_M$ will correspond exactly to the doubly-scattered light exiting the medium, which is known in closed form~\cite{sears75}
        \begin{equation}\label{psi2}
            f_2(\ppdir_i,\ppdir_o) = \frac{\ssalbedo^2}{4 \pi} \frac{\ppdir_i \arccoth(1+2 \ppdir_i) + \ppdir_o \arccoth(1+2 \ppdir_o )}{\ppdir_i + \ppdir_o}.
        \end{equation}
        where the direction cosines are $\ppdir_i = \cos \theta_i$ and $\ppdir_o = \cos \theta_o$.
        The component of the BRDF due to the negative mirrored uncollided term is
        \begin{multline}
            f_{un-}(\ppdir_i,\ppdir_o,z_b) = \frac{\ssalbedo^2}{4 \pi} \iint \e^{-u} \e^{-v} \phiun(v \ppdir_o,-u \ppdir_i - 2 z_b) du dv = \\
            \frac{- \ssalbedo^2 \text{Ei}(-2 z_b)}{4 \pi} + \frac{\alpha ^2 \left(\ppdir_i
               e^{\frac{2 z_b}{\ppdir_i}} \text{Ei}\left(\frac{2 (\ppdir_i+1)
               z_b}{-\ppdir_i}\right)-\ppdir_o e^{\frac{2 z_b}{\ppdir_o}}
               \text{Ei}\left(\frac{2 (\ppdir_o+1) z_b}{-\ppdir_o}\right)\right)}{4
               \pi  (\ppdir_i-\ppdir_o)}.
        \end{multline}
        where we have used $\eta = 1$ to ignore refraction for now.
        A similar analysis gives the positive and negative BRDF terms due to the diffusive sources:
        \begin{equation}
            f_{D+}(\ppdir_,\ppdir_o) = \frac{2 \pi  (2 \mueff \ppdir_i
   \ppdir_o+\ppdir_i+\ppdir_o)}{\mueff (\mueff
   \ppdir_i+1) (\mueff \ppdir_o+1)
   (\ppdir_i+\ppdir_o)}
        \end{equation}
        and
        \begin{equation}
            f_{D-}(\ppdir_,\ppdir_o,z_b) = -\frac{2 \pi  e^{-2 \mueff z_b}}{\left(\mueff^2
   \ppdir_i+\mueff\right) (\mueff \ppdir_o+1)}
        \end{equation}
        The associated BRDF for the multiply-scattered light $f_m$ is thus
        \begin{multline}\label{eq:myfr}
            f_m(\ppdir_i,\ppdir_o) = \ssalbedo^2 \, \Ccol \left[ f_{D+}(\ppdir_i,\ppdir_o) - a_D \, f_{D-}(\ppdir_i,\ppdir_o,z_{bD}) \right] \\ - a_{un} \, f_{un-}(\ppdir_i,\ppdir_o,z_{bun}) + f_2(\ppdir_i,\ppdir_o)
        \end{multline}




%% file: benchmarking.tex
\section{Benchmarking Angular BSSRDFs}\label{sec:bench}
  
  To optimize for mirror distances $z_{bun}$ for positioning the negative uncollided fluence sources, and $z_{bD}$ for positioning the negative diffusive fluence images (in either the BRDF or the BSSRDF form) and for the scaling weights $a_{un}$ for the negative uncollided term and $a_D$ for the negative diffusive term, we choose the exact solution for the half-space albedo problem~\cite{chandrasekhar60} as our accuracy benchmark:
    \begin{equation}
        f_r(\ppdir_i,\ppdir_o) = \frac{\ssalbedo}{4 \pi} \frac{H(\ppdir_i) H(\ppdir_o)}{\ppdir_i + \ppdir_o}
    \end{equation}
    thereby prioritizing accurate angular variation and total reflectances above low order scattering of isolated beam illumination (since piece-wise uniform/smooth lighting scenarios are far more prevelant).  This solution uses Chandrasekhar's H-function.  To compare to the associated BRDF we derived in the previous section, we must subtract the singly-scattered reflectance,
    \begin{equation}\label{psi1}
        f_1(\ppdir_i,\ppdir_o) = \frac{\ssalbedo}{4 \pi} \frac{1}{\ppdir_i + \ppdir_o}
    \end{equation}
    to form the multiple-scattering BRDF, $f_m = f_r - f_1$, which we compare to Equation~\ref{eq:myfr}.

\begin{figure}
        \centering
        \subfigure[]{\includegraphics[width=0.48 \linewidth]{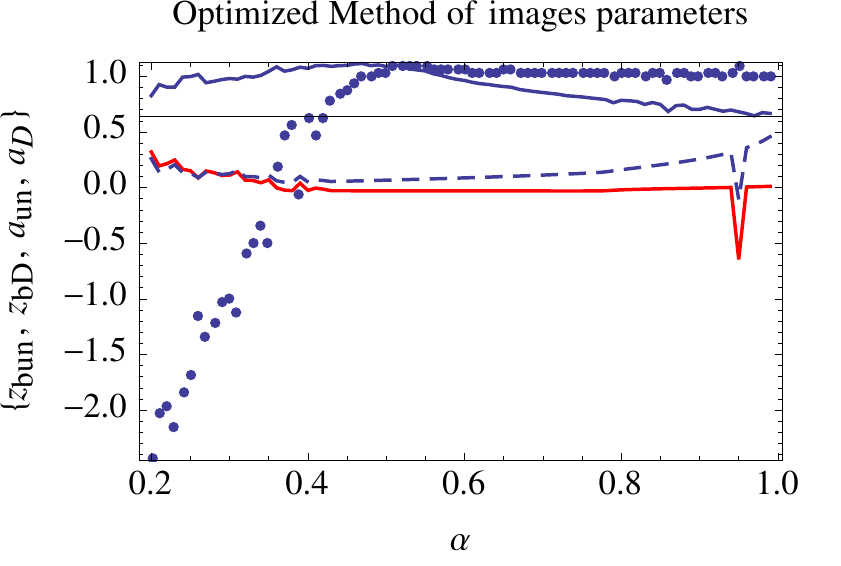}}
        \subfigure[]{\includegraphics[width=0.48 \linewidth]{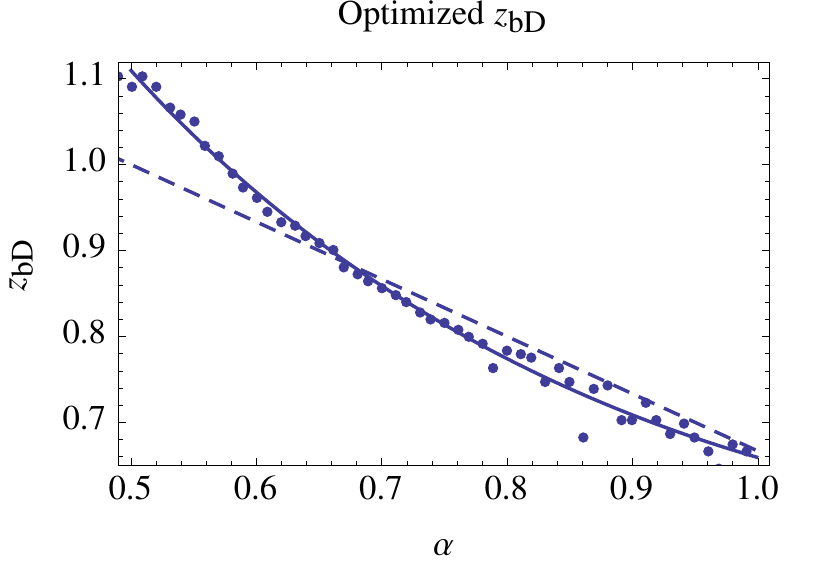}}
        \caption{a) Optimized method-of-images parameters for various single-scattering albedos, $\ssalbedo$.  The four curves are $z_{bun}$ (red), $z_{bD}$ (continuous blue), $a_{un}$ (dashed), and $a_D$ (circles). b) Comparison of the optimal diffusive extrapolation distance $z_{bD}$ (circles) and that predicted by diffusion theory $z_b = 2 D$ (dashed).  Our new optimized result (continuous curve) adapts to an optimal configuration not predicted by diffusion theory.}
        \label{fig-fourfits} 
    \end{figure}

    We fixed $\ppdir_i = 1$ to optimize for normally-incident illumination and sampled $f_m$ at $20$ discrete values for $\ppdir_o \in [0,1]$.  We performed a Levenberg-Marquardt optimization (using a mean square difference between the associated BRDF and the benchmark solution) to solve for $\{z_{bun}, z_{bD}, a_{un}, a_D\}$ given $\ssalbedo$.  The optimal coefficients are shown in Figure~\ref{fig-fourfits}a as a function of $\ssalbedo$.  We see that the optimal solution is similar to the Grosjean diffusion configuration proposed by d'Eon and Irving~\cite{deon11a}---for low absorption levels the optimal extrapolation distance $z_{bD}$ is close to $2 D$ (Figure~\ref{fig-fourfits}b) and the scale factor of the negative diffusive source is near $1.0$.  It is also clear that the negative uncollided term is useful for improving the accuracy of the method of images for the half space and its optimal extrapolation distance is near $0.0$ (but with an intensity that varies with absorption level and much less than $1.0$).

    \begin{figure*}
        \centering\parbox{0.7\linewidth}{%
    \setlength{\pwidth}{0.195\linewidth}%
    \setlength{\gapwidth}{0.01\linewidth}%
    \begin{rotate}{90}\parbox{0.5\pwidth}{\small\centering\textfig{MC}}\end{rotate}%
    \hfill%
    \includegraphics[width=\pwidth,angle=0]{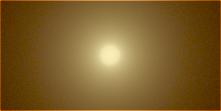}%
    \hfill%
    \includegraphics[width=\pwidth,angle=0]{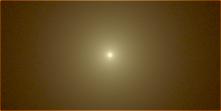}%
    \hfill%
    \includegraphics[width=\pwidth,angle=0]{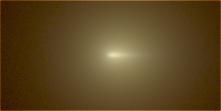}%
    \hfill%
    \includegraphics[width=\pwidth,angle=0]{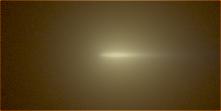}%
    \hfill%
    \includegraphics[width=\pwidth,angle=0]{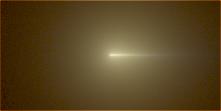}%
    \\[0.02\gapwidth]
    \begin{rotate}{90}\parbox{0.5\pwidth}{\small\centering\textfig{Our}}\end{rotate}%
    \hfill%
    \includegraphics[width=\pwidth,angle=0]{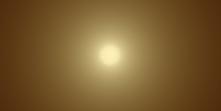}%
    \hfill%
    \includegraphics[width=\pwidth,angle=0]{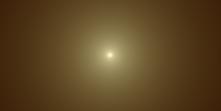}%
    \hfill%
    \includegraphics[width=\pwidth,angle=0]{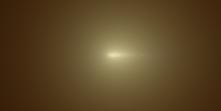}%
    \hfill%
    \includegraphics[width=\pwidth,angle=0]{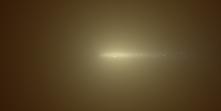}%
    \hfill%
    \includegraphics[width=\pwidth,angle=0]{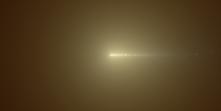}%
    \\[0.02\gapwidth]
    \begin{rotate}{90}\parbox{0.5\pwidth}{\small\centering\textfig{PBD}}\end{rotate}%
    \hfill%
    \includegraphics[width=\pwidth,angle=0]{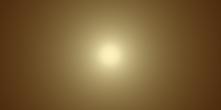}%
    \hfill%
    \includegraphics[width=\pwidth,angle=0]{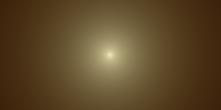}%
    \hfill%
    \includegraphics[width=\pwidth,angle=0]{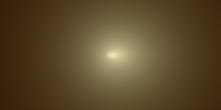}%
    \hfill%
    \includegraphics[width=\pwidth,angle=0]{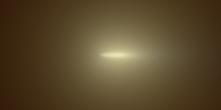}%
    \hfill%
    \includegraphics[width=\pwidth,angle=0]{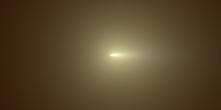}%
    \\[-0.2\gapwidth]
    \begin{rotate}{90}\parbox{\pwidth}{\small\centering\phantom{\textfig{dummy}}}\end{rotate}%
    \hfill%
    \parbox{\pwidth}{\small\centering\textfig{$r_b = 1.0, \theta_i = 0^\circ$}}%
    \hfill%
    \parbox{\pwidth}{\small\centering\textfig{$r_b = 0.2, \theta_i = 0^\circ$}}%
    \hfill%
    \parbox{\pwidth}{\small\centering\textfig{$r_b = 0.2, \theta_i = 60^\circ$}}%
    \hfill%
    \parbox{\pwidth}{\small\centering\textfig{$r_b = 0.2, \theta_i = 80^\circ$}}%
    \hfill%
    \parbox{\pwidth}{\small\centering\textfig{$r_b = 0.0, \theta_i = 80^\circ$}}%
  }%
  \caption{Reflected radiance along the outward normal due to multiple scattering of beam illumination striking a homogeneous half space with isotropic scattering and indexed-matched ($\eta = 1$) boundaries.  Our new dual-beam BSSRDF (middle row) closely matches Monte Carlo simulation (MC).  The recent Photon Beam Diffusion (PBD) model computes the total exitant flux, not the angularly-resolved exitant radiance, and thus is not reciprocal and lacks accurate total reflectances at oblique angles.  The angle of incidence $\theta_i$ and radius of the beam $r_b$ (in mean-free-paths) are varied in each column.  The mean free path of the medium is $1$ at all wavelengths, with spectral single-scattering albedo $\{0.99, 0.91, 0.5\}$.  An unrealistic gamma correction of $6.0$ was applied to the images to compare to [Habel et al. 2013].
          \label{fig:teaser}}
    \end{figure*}

\begin{figure*}
        \centering
        \subfigure[]{\includegraphics[width=0.33 \linewidth]{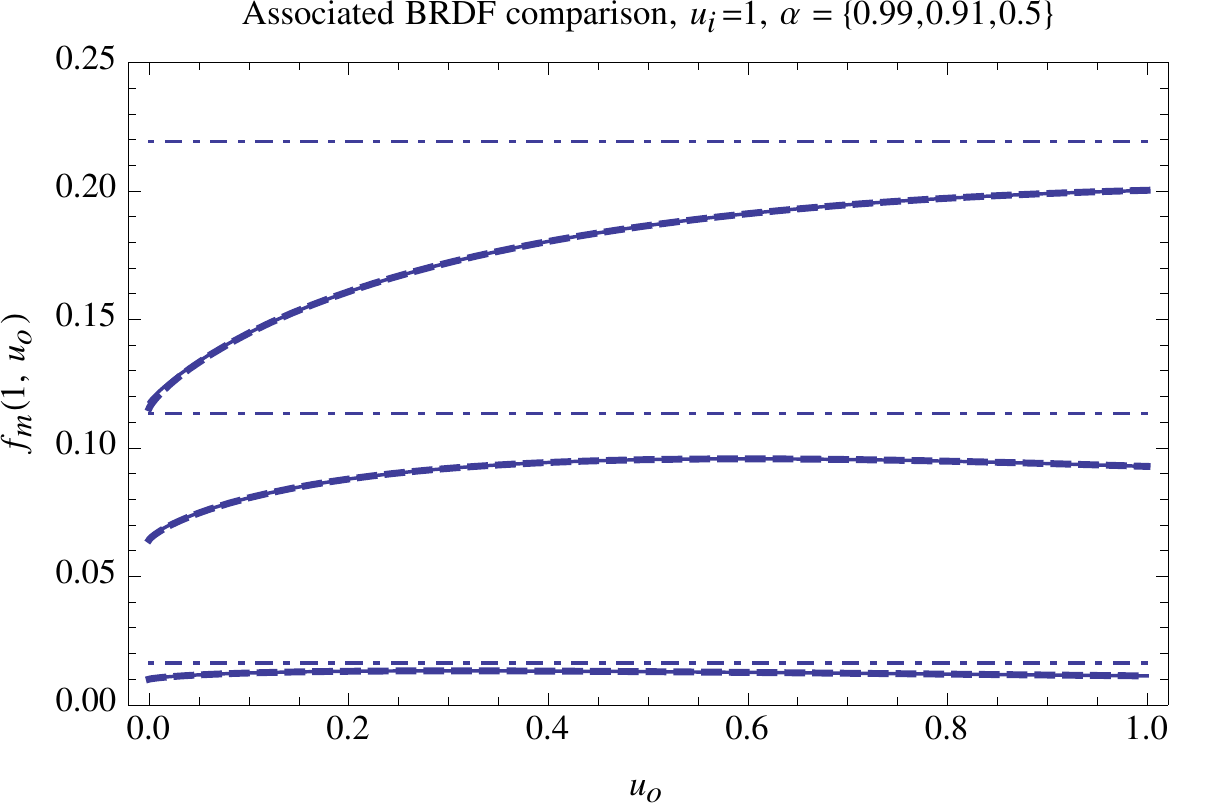}}
        \subfigure[]{\includegraphics[width=0.33 \linewidth]{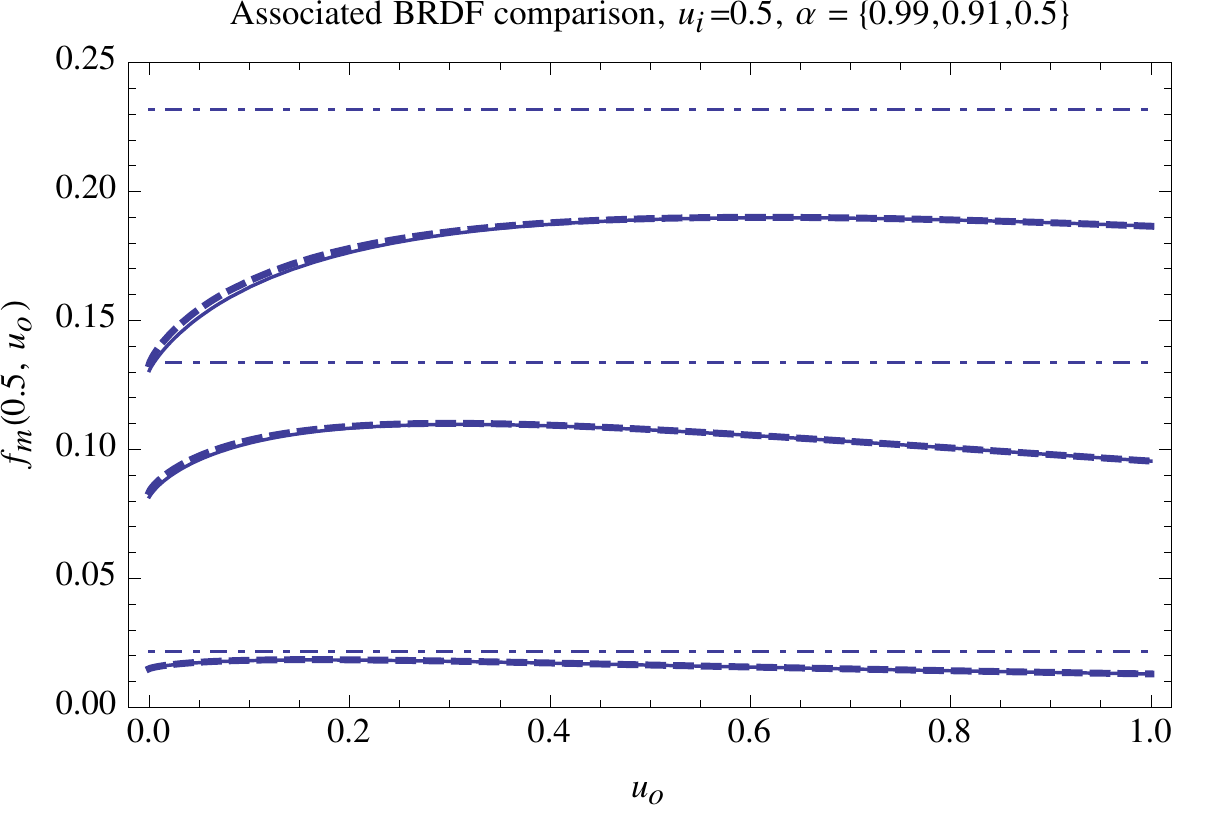}}
        \subfigure[]{\includegraphics[width=0.33 \linewidth]{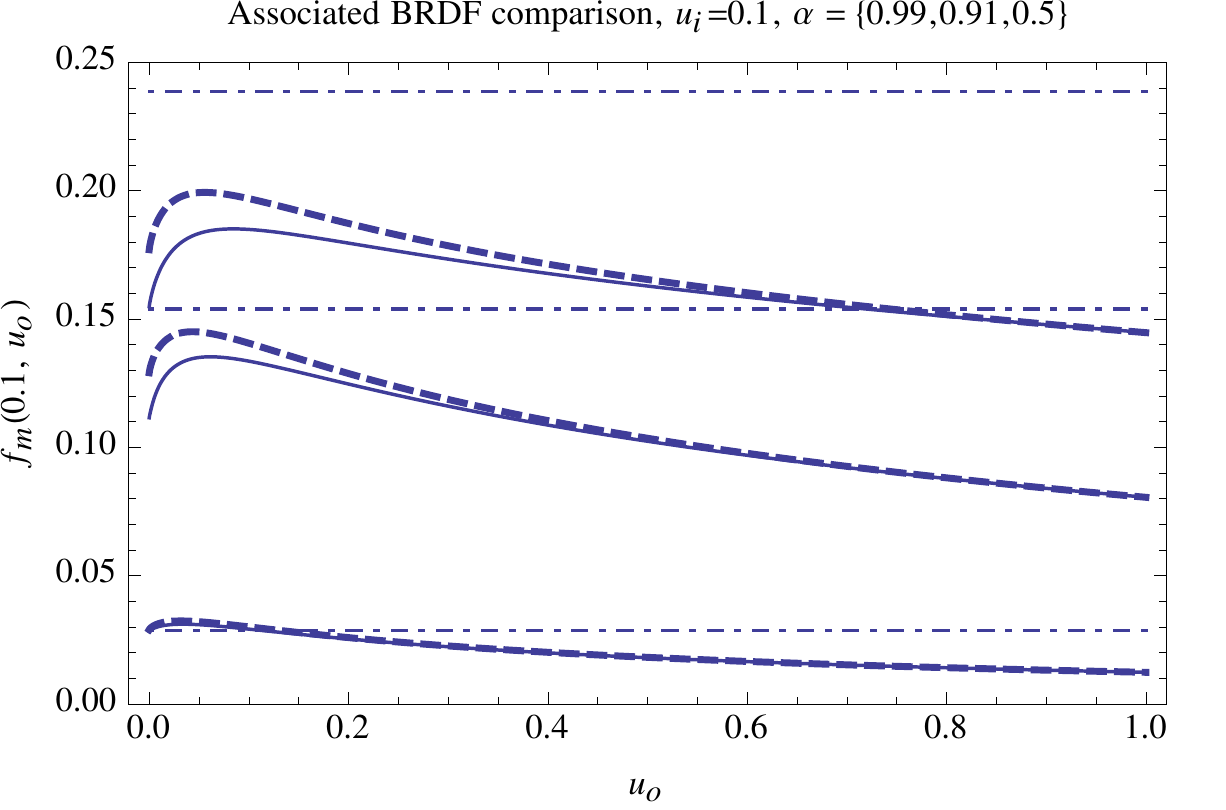}}
        \caption{The multiply-scattered portion $f_m$ of the BRDF for the half-space with isotropic scattering and indexed-matched boundary conditions as predicted by our dual-beam BSSRDF (continuous curve), Photon Beam Diffusion (dot-dashed) and the exact solution (thick-dashed).  The Photon Beam Diffusion method enforces a diffuse angular exitance shape and does not accurately predict the variation in intensity of a uniformly illuminated surface as the angle of view changes.}
        \label{fig-BRDF} 
    \end{figure*}

    In Figure~\ref{fig-BRDF} we illustrate the behaviour of the optimized dual-beam BRDF for three absorption levels $\ssalbedo \in \{0.99, 0.91, 0.5\}$ and for three incident angles.  In all cases the dual-beam method-of-images BRDF is quite close to the benchmark solution and exhibits desired angular and total reflectance accuracy that Photon Beam Diffusion~\cite{habel13} lacks.  The optimized parameters for $\ssalbedo = 0.99$ were $z_{bun} = 0.011, z_{bD} = 0.667, a_{un} = 0.457, a_D = 1.01$, for $\ssalbedo = 0.91$ were $z_{bun} = -0.003, z_{bD} = 0.697, a_{un} = 0.27, a_D = 1.$, and for $\ssalbedo = 0.5$ were $z_{bun} = -0.0285, z_{bD} = 1.089, a_{un} = 0.0671, a_D = 1.036$.  An approximate fit for the mirror distances (in mean free paths) and scale factors for $\ssalbedo > 0.5$ is:
    \begin{align*}
      z_{bD}(\ssalbedo) &= 0.335867 \ssalbedo^2-0.62166 \ssalbedo+\frac{0.944945}{\sqrt{\ssalbedo}} \\
      z_{un}(\ssalbedo) &= \text{max}(-0.03,0.154352 \ssalbedo-0.142497) \\
      a_{un}(\ssalbedo) &= -7.7\, +9.8 \ssalbedo^3-22.8 \ssalbedo^2+20 \ssalbedo+\frac{1.1}{\ssalbedo} \\
      a_D(\ssalbedo) &= 0.359563 \ssalbedo^2-0.692592 \ssalbedo+1.34954
    \end{align*}

\section{Application of the Dual-Beam BSSRDF}
  Given a BSSRDF that has been optimized to have accurate BDRF behaviour, we return to its application to render translucent materials.  Applying the BSSRDF requires evaluating both of the ``beam" integrals efficiently.  This is closely related to the classic problem of estimating flux at a point and indeed each integral can be regarded as a track-length estimator of the associated Green's function term.  For the uncollided terms, the singularity is of order $1/r^2$ and the equi-angular transform of Rief et al.~\shortcite{rief84} is useful for forming an accurate low-order quadrature.  We found that deriving a similar $1/r$ singularity transform for the diffusive terms, and applying a Gaussian quadrature post transformation allows a low number of regular samples to efficiently evaluate the track-length integrals.  These ideas are closely related to the application of track-length estimators in graphics~\cite{novak12,habel13}, and our evaluation of the dual-beam integration follows Novak for the most part.  To avoid quadrature errors from skewing the comparison in Figure~\ref{fig:teaser} we used 5000 exponential sample pairs to evaluate the dual-beam integral, where we show a beam of various widths striking a surface at various angles with spectral absorption levels.  The PBD result is discernibly more blurred and does not resolve the subsurface beam correctly, although the overall BRDF characteristics of our dual-beam BSSRDF seems to remain its larger strength, since such localized beam illumination is incredibly uncommon in typical image synthesis, whilst accurate colors are important for predictive rendering and level-of-detail transitions.
\section{Conclusions and Future Work}
  We have presented a novel mathematical framework for designing method-of-images 3D searchlight BSSRDFs.  This was used to derive a new dual-beam BSSRDF for an isotropic half space that closely matches the benchmark BRDF solution.  The result is a BRDF/BSSRDF pair that can be readily applied using off the shelf methods for beam-integrals, and that is also reciprocal and produces accurate colors and angular variation with oblique illumination and viewing angles.  Future work involves a more thorough investigation of image configurations for boundaries with smooth~\cite{williams06} and rough Fresnel interfaces.  The latter can be readibly applied by extending Davison's line-integral of the exitant radiance stochastically by sampling the rough BTDF for the interface (and similarily for the source beam).  The novel method-of-images BRDFs may also be of use on their own for approximating coupled specular/subsurface behaviours.